\documentclass[%
reprint,
showpacs,
 amsmath,amssymb,
 prl,
]{revtex4-1}

\usepackage{graphicx}
\usepackage{dcolumn}
\usepackage{bm}

\usepackage{dcolumn}
\usepackage{bm}
\usepackage{nicefrac}
\usepackage{array}
\usepackage{amssymb}
\usepackage{amsmath}
\usepackage{graphicx}
\usepackage{multirow}
\usepackage{comment}
\usepackage[normalem]{ulem}
\usepackage{url}
\usepackage{mathrsfs}
\usepackage{float}
\usepackage{tabularx}
\usepackage[dvipsnames]{xcolor}

\newcommand{\rev}[1]{{\color{black}{#1}}}

\begin{document}
\pacs{82.20.Xr, 82.37.Np, 71.15.Mb, 68.43.Fg}
\keywords{tunneling, instanton, density functional theory, porphycene, double hydrogen transfer}

\title{Multidimensional Hydrogen Tunneling in Supported Molecular Switches: \\ The Role of Surface Interactions}

\author{Yair Litman}
\email{litman@fhi-berlin.mpg.de}
\affiliation{%
 Fritz Haber Institute of the Max Planck Society, Faradayweg 4--6, 14195 Berlin, Germany}%
\affiliation{ 
 Institute for Chemistry and Biochemistry, Freie Universit\"at Berlin, Arnimallee 22, 14195 Berlin, Germany}
\author{Mariana Rossi}%
\email{mariana.rossi@mpsd.mpg.de}
\affiliation{%
 Fritz Haber Institute of the Max Planck Society, Faradayweg 4--6, 14195 Berlin, Germany}%
\affiliation{MPI for the Structure and Dynamics of Matter, Luruper Chaussee 149, 22761 Hamburg, Germany}

\date{\today}

\begin{abstract}

The nuclear tunneling crossover temperature ($T_c$) of hydrogen transfer reactions in supported molecular-switch architectures can lie close to room temperature. This calls for the inclusion of nuclear quantum effects (NQE) in the calculation of reaction rates even at high temperatures. 
However, computations of NQE relying on \rev{standard} parametrized dimensionality-reduced models quickly become inadequate in these environments. 
In this letter, we study the paradigmatic molecular switch based on porphycene molecules adsorbed on metallic surfaces with full-dimensional calculations that combine density-functional theory for the electrons with the semi-classical ring-polymer instanton approximation for the nuclei. We show that the double intramolecular hydrogen transfer (DHT) rate can be enhanced by orders of magnitude due to surface fluctuations in the deep tunneling regime. We also explain the origin of an Arrhenius temperature-dependence of the rate below $T_c$ and why this dependence differs at different surfaces. 
We propose a simple model to rationalize the temperature dependence of DHT rates spanning diverse fcc [110] surfaces.
\end{abstract}

\maketitle

Nuclear tunneling is an inherently quantum-mechanical process that can strongly impact the properties of matter in a wide variety of situations, ranging from biological enzymes to organic-based technologies~\cite{LayfieldHammes-Schiffer2013, JiangShuai2015, KaapLeeuw2016, Koch_JACS_2017}. In complex environments, it has been recognized that signatures of tunneling on rate processes are often not well captured by textbook theories~\cite{Meisner_Angew_2016,Topaler_JCP_1994} and in particular, in hydrogen transfer reactions, the small mass of hydrogen makes tunneling pronounced~\cite{HYDROGEN_TRANSFER}. Still, a theoretical description of nuclear tunneling that goes beyond a simple one-dimensional approximation\cite{Gerritzen_JACS_1984,Limbach_2006} and considers anharmonic coupling between many degrees of freedom in larger-scale systems remains a challenge~\cite{Fang_NatCom_2020, Rommel_JPCB_2012,Mills_CPL_1997}.
To build up a systematic understanding of multidimensional rate processes in the deep tunneling regime and the impact of the environment on hydrogen dynamics, a fully \textit{ab initio} treatment of simple, yet non-trivial, reactions in well controlled conditions is desired.

\begin{figure}[htbp]
    \centering
    \includegraphics[width=0.70\columnwidth]{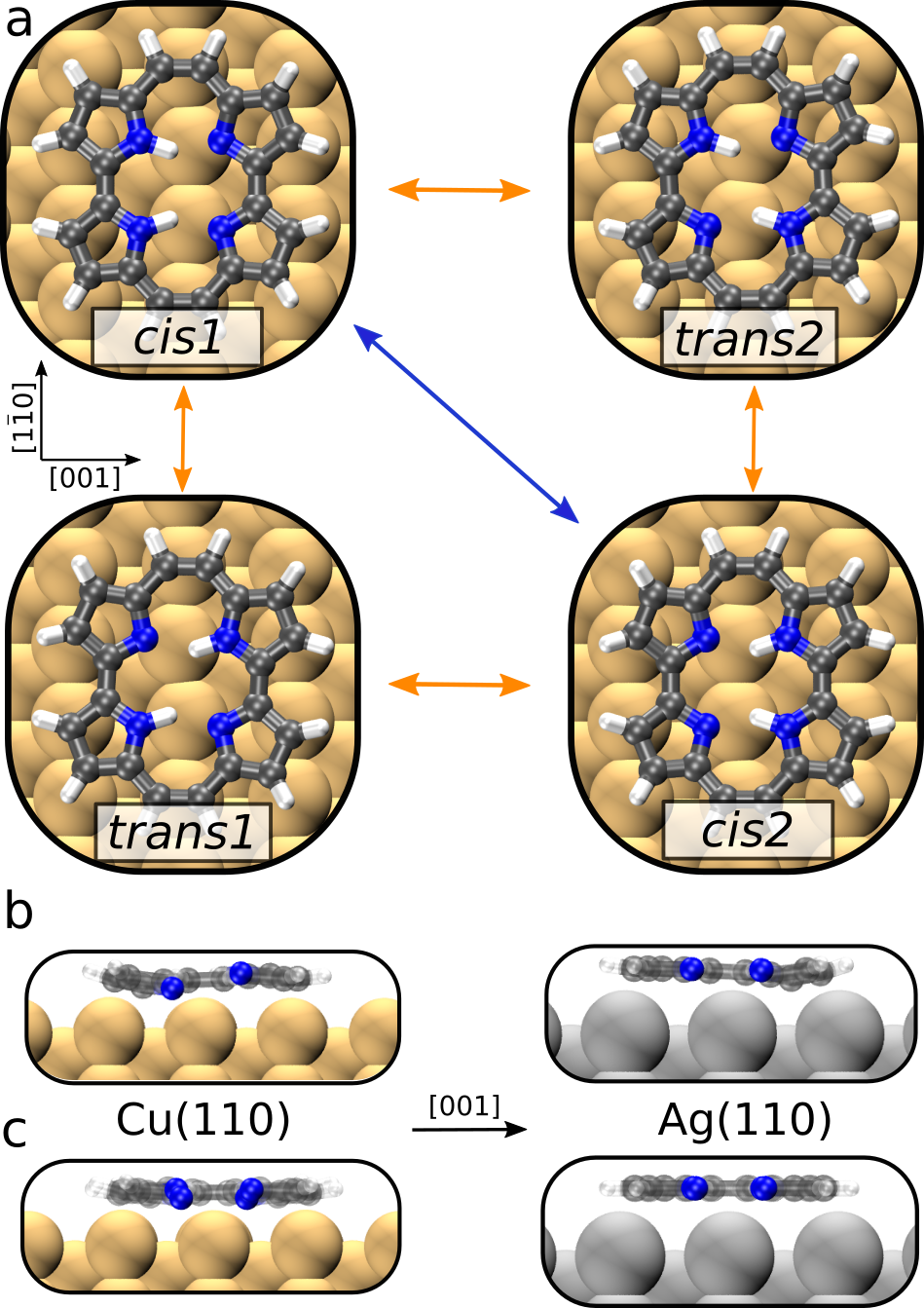}
    \caption{(a) Top view of the local minima of porphycene on Cu(110). Concerted and stepwise DHT mechanisms are represented by blue and orange arrows, respectively. Lateral view of the \textit{cis} (b) and \textit{trans} (c) conformers, of porphycene on Cu(110) and Ag(110).}  
    \label{fig:Porphycene}
\end{figure}

 Tetrapyrrole macrocycles, like porphyrin, naphtalocianine, and porphycene, show a remarkable diversity of functional properties~\cite{Auwarter_NatChem_2015, Kobke_NatNano_2020,Seufert_NatChem_2011,Shubina_JACS_2007,Grill_Rev_2008}, and were proposed, among others, as molecular switches~\cite{Zhang_ChemSocRev_2015,Waluk_ChemRev_2017,Qiu_PRL_2004}. The intramolecular hydrogen transfer reaction that occurs in the inner cage of these molecules, known as tautomerization, can be triggered remotely by different external stimuli~\cite{Liljeroth_2007,Kugel_NanoLett_2017,Kumagai_PRL_2013,Bockmann_NanoLett_2016,Ladenthin_NAT_2016,Mangel_PCCP_2020}. Moreover, because the reaction takes place without a pronounced conformational change, these molecules can be incorporated in nanoscale devices. Within this area, controlling the dynamical properties of these molecules is central to advance rational design. Molecular diffusion and rotations have been more often addressed~\cite{Eichberger_NanoLett_2008,Marbach_ChemComm_2014,Shea_JCP_2014,Buchner_JPCC_2011,Sonnleitner_PRL_2011}, while key aspects of the hydrogen transfer reaction mechanism and its temperature dependence remain poorly understood. This is especially due to the challenges in the description of multidimensional quantum dynamics that cannot be captured by perturbative treatments of anharmonic couplings~\cite{Warshel_1982_JPC, HabershonMano_Rev_2013, Wei_IntRev_2019, LitmanRossi2019,Meisner_Angew_2016}.

In this Letter, 
we study the effect of tunneling on the double intramolecular hydrogen transfer (DHT) of two representative systems, namely porphycene on Cu(110) and Ag(110) surfaces. In these reactions, the tunneling crossover temperature, which represents the temperature below which tunneling becomes greater than classical hopping over the barrier ($T_\text{c}=\hbar \omega_b /2\pi k_\text{B}$, where $\omega_b$ is the imaginary frequency of the unstable mode at the transition state geometry \cite{Gillan1987}) lies close to room temperature.
Specifically, we i) identify the multi-dimensional DHT mechanisms at different temperatures, ii) clarify their temperature dependence, and iii) elucidate the role of surface fluctuations in the deep tunneling regime. The results we obtain are able to explain puzzling experimental measurements \cite{Kumagai_PRL_2013,Koch_JACS_2017} that showed an unexpected temperature dependence of the DHT. In addition, they show that instead of acting only as a passive observer of the reaction, in certain situations the surface takes a prominent role in the tunneling event.

We here employ a combination of density-functional theory (DFT) for the electronic degrees of freedom with the ring polymer instanton (RPI) approximation~\cite{JOR_JCP_2009, Arni_thesis} for the nuclear degrees of freedom. 
RPI can be viewed as an extension of Eyring~\cite{Eyring} transition state theory (TST) which includes tunnelling and captures anharmonic contributions along the reaction pathway. 
It is a semi-classical method that uses discretized closed Feynman path (CFP) integrals to evaluate tunneling rates in the deep tunneling regime. 
This approximation finds the dominant stationary pathways in the CFP that connect reactants and products, the instanton pathways. Then, a series of steepest descent approximations to evaluate the flux-side correlation function leads to the instanton approximation for thermal reaction rates $k_\text{inst}$~\cite{Jor_review_2018}. The rate can be expressed as
\begin{equation} \label{eq:instanton}
\begin{split}
k_\text{inst}(\beta) &= A_\text{inst}(\beta) e^{-S[\mathbf{x}_\text{inst}(\beta)]/\hbar},
\end{split}
\end{equation}
where  $A_\text{inst}$ accounts for the harmonic fluctuations around the instanton pathway $\mathbf{x}_\text{inst}$, $S$ is the Euclidean action, and $\beta=1/k_\text{B}T$ with $k_\text{B}$ the Boltzmann constant and $T$ the temperature. In the discretized CFP space, $\mathbf{x}_\text{inst}$ is a first order saddle point. Albeit approximate, this method shows the best tradeoff in situations where the quantum exponential wall would prevent a full dimensional evaluation of the exact tunneling rate \cite{Richardson_PCCP_2017,Topaler_JCP_1994}. 

Within the Born-Oppenheimer (BO) approximation, evaluating Eq.~\ref{eq:instanton} requires the BO energies and forces. We here employ the most accurate level of theory affordable for the system size and number of system replicas required. We employ DFT with the Perdew-Burke-Ernzerhof (PBE) exchange-correlation functional~\cite{PBE} including the  Tkatchenko-Scheffler~\cite{TS} dispersion correction modified to treat physisorption on surfaces~\cite{TSsurf}.
We thereby ensure a good description of the anisotropic electron density redistribution and the interfacial orbital coupling that takes place upon porphycene adsorption~\cite{Kumagai_JCP_2018, Jingtai_JCPL_2019}. At this level of theory, we do not expect reaction barriers to be quantitatively accurate~\cite{Barone_CPL_1994}, but expect to capture qualitative trends and the correct physics. In the supplemental material (SM), we report selected geometries calculated with a range-separated hybrid functional for comparison and discuss the importance of vdW dispersion interactions in this context \cite{Stohr_ChemSocRev_2019, Guirong_ProgSurfSci_2019}.
These calculations were enabled by the combination of the FHI-aims \cite{FHI-AIMS} all-electron code and the i-PI \cite{i-pi,i-pi2} universal force engine. 
Details and convergence tests are provided in the SM which also includes references 
\cite{ASE, MBD-nl,litman_thesis,Litman_FD_2019,Zhang_PCCP_2014,Garg_AmPhys_2000,Litman_JCP_2018,Eyring_1935_JCP,HSE06_mod}.

\begin{figure*}[htbp]
       \includegraphics[width=0.9\textwidth]{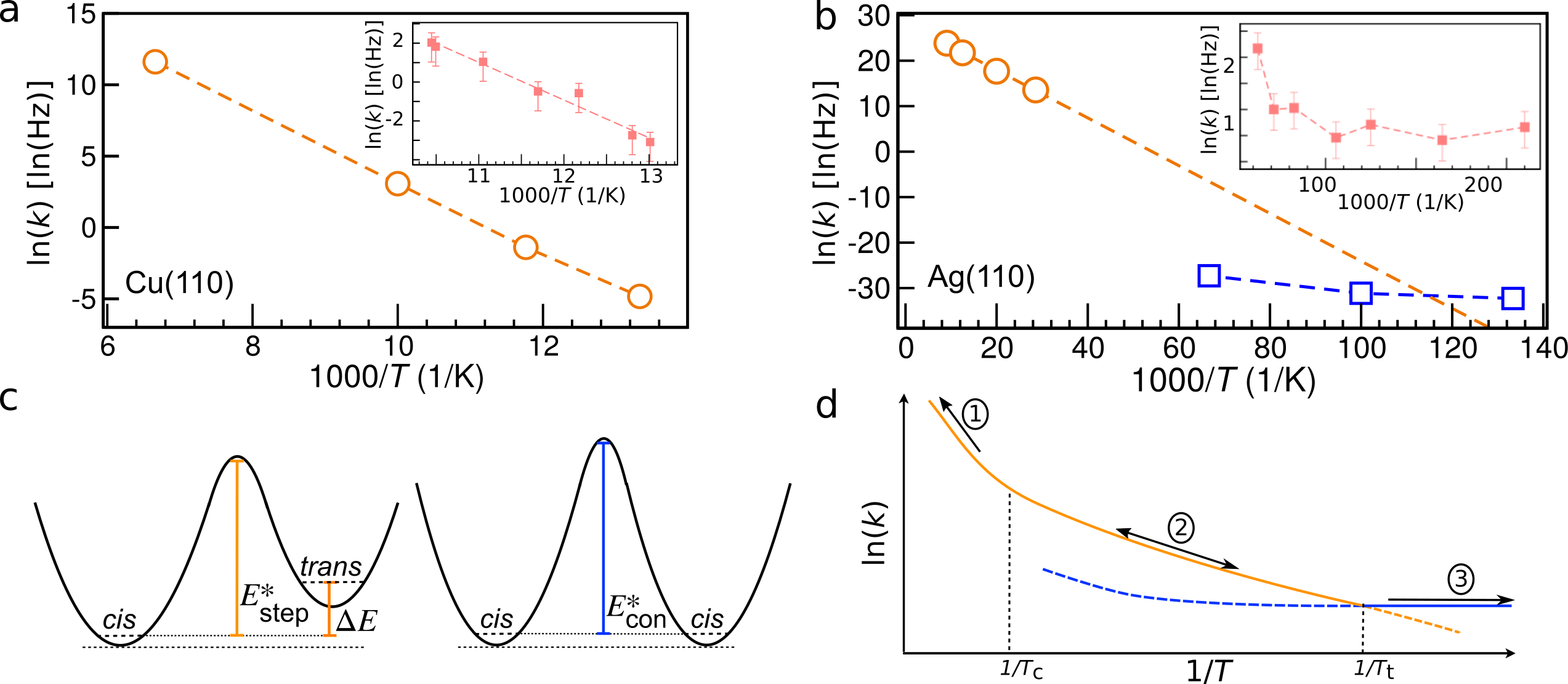}
    \caption{(a) Calculated DHT rates $k_{\text{inst}}$ for the  \textit{cis} $\rightarrow$ \textit{cis} reaction of porphycene on Cu(110) between 75 and 150 K. At these temperatures, $k_{\text{inst}}$ only includes contributions from the stepwise mechanism . The inset shows experimental results from Ref.~\cite{Kumagai_PRL_2013}.
    (b) Calculated DHT $k_{\text{inst}}$ for the  \textit{cis} $\rightarrow$ \textit{cis} reaction of porphycene on Ag(110) between 7.5 and 110 K. Orange circles represent  $k_{\text{inst}}$ of the stepwise mechanism and blue squares of the concerted mechanism. The inset shows experimental results from Ref.~\cite{Koch_JACS_2017}.
    (c) Schematic 1D potential energy surfaces of the stepwise (left) and concerted (right) reactions. The ZPE for reactant and products are shown, while the ZPE at the barrier top is not (but it was included in the multidimensional calculations).
    (d) Schematic representation of the different temperature dependence regimes of the stepwise (orange) and concerted (blue) DHT rate of adsorbed porphycene. Full/dashed lines represent the dominant/minor mechanism.
    The numbers 1, 2 and 3 illustrate the three different regimes, see text.}  
    \label{fig:Cu110Rates}
\end{figure*}

In porphycene, the inner-cage hydrogen atoms can adopt different configurations, giving rise to two stable tautomers, coined \textit{cis} and \textit{trans}. Each tautomer is two-fold degenerate and their structures on Cu(110) and Ag(110) are shown in Fig.~\ref{fig:Porphycene}. On these surfaces, the {\it cis} tautomer is more stable than the {\it trans} tautomer by 172 and 20 meV, respectively. Indeed, on the same surfaces, experiments observe almost exclusively the \textit{cis} tautomer~\cite{Ladenthin_NAT_2016,Koch_JACS_2017,Kumagai_NatCom_2014}. The DHT of the \textit{cis}$\rightarrow$\textit{cis} reaction can happen through two possible mechanisms. The concerted mechanism, where the hydrogen atoms are transferred together, without the existence of a stable intermediate and passing through a second-order saddle point, and the stepwise mechanism, where the hydrogen atoms are transferred sequentially and the reaction involves a \textit{trans} intermediate. 

We have recently studied the DHT of porphycene in the gas phase~\cite{LitmanRossi2019}, and found that there is a competition between a concerted and a stepwise DHT at different temperatures.
However, new effects arise due to the interaction between the molecule and the surface.  
In order to rationalize them, we divide the influence of the surface on the physisorbed molecule into static and dynamical effects. The static surface effects refer to the change in the potential energy landscape upon adsorption, and can affect both static and dynamical properties of the adsorbed molecule. For example, while the global minimum in the gas phase is the \textit{trans} tautomer in a flat conformation, on Cu(110) and Ag(110) the relative tautomer stability is reversed and the molecule is buckled, as shown in Fig. \ref{fig:Porphycene}b and \ref{fig:Porphycene}c. This distortion changes the DHT energy barriers and consequently the hydrogen dynamics. 
Dynamical surface effects, on the other hand, refer to the
impact that the motion of the surface atoms have on the molecular properties. As it will be shown, they can significantly influence the dynamics of the DHT.

On Cu(110), we calculate $T_c = 264$ K for the DHT and show the instanton tunneling rates of the \textit{cis}$\rightarrow$\textit{cis} reaction in Fig.~\ref{fig:Cu110Rates}a.
We show rates between 75 and 150 K because they lie well within the deep tunneling regime and we wish to compare with experimental results in the same temperature range~\cite{Kumagai_PRL_2013}, shown in the inset. In this case, we find a very good agreement with experiment throughout the six-orders of magnitude variation of the rate. 
The calculated effective activation energy $E_\text{A}$ is 190 meV.  This compares well with the value of 168 $\pm$ 12 meV reported by experiments. 
Both the calculated rates and the measured ones show an Arrhenius-like temperature dependence even though tunneling is dominant at this temperature range. We proceed to explain the origin of this dependence.

The calculated DHT rates contain contributions almost exclusively from the stepwise mechanism, because at these temperatures it is several orders of magnitude faster than the concerted alternative. We propose that the  \textit{trans} intermediate was not observed because its 
predicted residence time is $\approx$ 0.1 ns, which lies beyond the time resolution of STM experiments ($\approx 100$ $\mu \text{s}$). 
The instanton trajectory provides an intuitive view of the reactive process by showing the main ``instantaneous" tunneling configuration that the delocalized nuclei adopt. The trajectory is visualized in Fig. S1 at several temperatures. As the temperature is decreased, the reaction takes shorter pathways and crosses regions of higher PES energy. Moreover, we observe a considerable contribution from heavier atoms like C and N and, interestingly, even Cu atoms to the tunneling mechanism (see Tab. S11). These are manifestations of the multidimensional nature of the tunneling process\rev{~\cite{Tuckerman_PRL_2001,Chenfang_2019_JPCL,Fang_NatCom_2020}}. \rev{They show} that reducing the problem dimensionality, as previously done for porphyrin and other 
cyclic hydrogen bonded solids through the Bell-Limbach model \cite{Limbach_2006,Klein_JACS_2004}, \rev{without at least considering a temperature dependence of the parameters,} is inappropriate in this case.

After these considerations, 
the observed $E_\text{A}$ can be understood as follows. At low enough temperatures, when only the vibrational ground states (VGS) are populated and a further decrease of the temperature does not affect the vibrational populations, the \textit{trans}$\rightarrow$\textit{cis} reaction will proceed from the VGS of the reactant and will be constant with temperature.
As a consequence and because of detailed balance, the inverse reaction \textit{cis}$\rightarrow$\textit{trans}, which is the rate-controlling step of the stepwise mechanism, must show $E_\text{A}$ equal to the difference between the VGS energies of reactant and product, that we call $\Delta E$\rev{~\cite{Limbach_2006}} (see Fig. \ref{fig:Cu110Rates}c). The height and width of the barrier impact the absolute value of the rate, but they do not affect the Arrhenius slope in the low temperature limit of an asymmetric reaction.
Indeed, a harmonic estimation of $\Delta E$ is 172 meV, which is very close to the calculated $E_A=190$ meV from Fig. \ref{fig:Cu110Rates}a. 

\begin{table}
\centering
	\begin{tabular}{c|c|c|c|c|c}
	\hline
	Surface &$T$ (K)& $\kappa_{\text{tun}}$ &  KIE$_{\text{inst}}$ & KIE$_{\text{TST}}$ & SFE \\
	\hline
	\hline
    Cu(110) & 100    &1.0      &     21               &     89     &   34  \\
    Cu(110) & 85     &1.2      &     32               &     197    &  106  \\
    Cu(110) & 75     &2.4      &     66               &     397    &  464  \\
    \hline
    Ag(110) & 75     &   14.9             &       -               &   -         &    4  \\
    \hline
	\end{tabular}
	\caption{Tunneling enhancement factor ($\kappa_{\text{tun}}$), 
	and kinetic isotopic effects (KIE) and surface fluctuations enhancement (SFE). See definitions and discussion in the text.}
	\label{tab:KIE} 
\end{table}

We further analyze how tunneling manifests itself in these reaction rates in Table \ref{tab:KIE}.
A standard procedure to estimate the impact of tunneling is to compare $k_{\text{inst}}$ with the rate predicted by the Eyring TST ($k_\text{TST}$), since the latter neglects tunneling but includes ZPE. The tunnelling enhancement factor is $\kappa_\text{tun} = k_\text{inst}/k_\text{TST}$ \cite{Beyer_PCL_2016}, which is reported in Tab.~\ref{tab:KIE}.
Surprisingly, because these factors are close to 1, they would seem to indicate that tunnelling plays a minor role. To understand this observation, we computed the  kinetic isotope effect (KIE), defined as $k^\text{H}/k^\text{D}$, where $k^\text{D}$ was obtained from calculations where the inner-cage hydrogen atoms were replaced by deuterium. 
If tunnelling would be a minor effect, the only difference in these rates should be ZPE, and since ZPE is captured by both $k_{\text{inst}}$ and $k_\text{TST}$, the KIE of both should be similar. However, as shown in Tab. \ref{tab:KIE}, these numbers are different. TST overestimates the KIE in this particular case, for reasons outlined in the SM, Section VII. 
We thus conclude that $\kappa_{\text{tun}} \approx 1$ because of the following.
On Cu(110) the $\Delta E$ (including harmonic ZPE) between reactants and products, which is a good estimate for $E_A$ in $k_\text{inst}$ as discussed above, happens to be similar to the energy difference between the ZPE corrected reactant and transition state, which defines $E_A$ for $k_\text{TST}$ (see Table \ref{tab:Surfaces}). Additionally, because close to $T_c$ both rates are comparable, if $E_A$ are similar, the prefactors also must be. This observation explains why TST fared reasonably well in these systems in the past, even without including the relevant physics of tunneling \cite{Jingtai_JCPL_2019}.
Thus, we propose that KIE$_\text{inst}$/KIE$_\text{TST}$ can be an alternative measure of tunneling contributions to hydrogen transfer reactions.

The investigation of the dynamical surface effects on the DHT
required RPI calculations where we fixed the surface atoms at the reactant position. The rates obtained as a result of this constrained optimization lack all contributions from  
fluctuations of the surface degrees-of-freedom. We call the ratio between the rates with and without those constraints the ``surface fluctuations enhancement" (SFE). Further details can be found in the SM.
The SFE for $k_\text{inst}$ on Cu(110) are reported in Table~\ref{tab:KIE} and can adopt surprisingly large values. The results show that the dynamical surface effects act on the opposite direction of the static ones, increasing the tunneling rates up to two orders of magnitude. 
Interestingly, the SFE become larger at lower temperatures because the contribution of heavy atoms to tunneling increases with decreasing temperature.

We then compare the Cu(110) with the Ag(110) substrate, a surface with a weaker static interaction. On Ag(110), porphycene is less buckled upon adsorption (see Fig.~\ref{fig:Porphycene}b) and the \textit{trans} conformer lies only 20 meV above \textit{cis}, but $T_c$ is also 264 K. Accordingly, it was observed in experiments that the DHT rates are substantially faster on Ag(110) than on Cu(110)~\cite{Koch_JACS_2017}.
Unlike Cu(110), the measured rates show two distinct regimes~\cite{Koch_JACS_2017}, reproduced in the inset of Fig.~\ref{fig:Cu110Rates}b. 
Above $\sim 10$ K there is an Arrhenius behaviour, while below $\sim 10$ K the rate shows almost no temperature dependence.
In Fig. \ref{fig:Cu110Rates}b we show the calculated $k_\text{inst}$ for the stepwise and the concerted mechanisms of the \textit{cis}$\to$\textit{cis} DHT of porphycene at Ag(110) (see calculation details in the SM).
We obtain an $E_\text{A}$ of 45 meV for the stepwise mechanism, which compares reasonably well with the experimental $E_A$, which we estimate to be 12 $\pm$ 3 meV. 
The calculated harmonic value of $\Delta E$ is again close to $E_\text{A}$ for this reaction.
At 10 K the $\mathbf{x}_\text{inst}$ for the concerted mechanism starts at the reactant minimum, indicating that tunneling takes place from the reactant VGS.
As such, the rates for the concerted mechanism (which is symmetric) do not change with temperature below this point and it becomes dominant below 8.5 K. Hence, we can explain the two regimes observed in experiment by the change in the DHT mechanism. The lack of quantitative agreement between the calculated and measured, transition temperature and rate of the concerted mechanism comes most likely from the potential energy surface used in the calculations, but some dependence of the measured rates on the STM tip cannot be fully discarded. Finally, as shown in Table
~\ref{tab:KIE}, the SFE are smaller here than they were for Cu(110), accounting for a factor 4 increase of the rate at 75 K. This is consistent with the weaker adsorbate-surface interaction strength.

Building up on these considerations, the dependence of these DHT rates can be schematically understood as shown in Fig.~\ref{fig:Cu110Rates}d and predicted for other metallic fcc [110] surfaces. One needs to compute the ZPE-corrected energy difference between the \textit{cis} and \textit{trans} tautomers $\Delta E$, the ZPE-corrected energy barrier for the stepwise $E^*_\text{step}$ and the $T_c$ of the stepwise reaction.
At high temperatures the reaction behaves classically and
proceeds by hopping over the lower barrier, which is normally the stepwise one, yielding a slope of $\approx E^*_\text{step}$, labeled 1 in Fig.~\ref{fig:Cu110Rates}d.
Considerably below $T_c$, the low temperature limit of the stepwise tunneling reaction is achieved and a slope of $\approx \Delta E$ should be observed (labeled 2). Finally, below a transition temperature $T_t$, the concerted mechanism becomes dominant and the rate becomes independent on temperature (labeled 3).

\begin{table}
	\begin{tabular}{c|c|c|c|c}
	\hline
	Surface  & $\Delta E$  (meV) & $T_c$ (K) & $E^*_\text{step}$ (meV)& $T_\text{t}$  (K)\\
	\hline
	\hline
    Ag(110)  &  20    &   264         &   61            &    3 - 9\\
    Au(110)  &  59    &   264         &   85            &   8  - 22 \\
    Cu(110)  &  172   &   264         &   222           &   15 - 41 \\
    Ni(110)  &  199   &   297         &   347           &   13 - 35 \\ 
    Pd(110)  &  148   &   293         &   326           &   12 - 34 \\ 
    \hline
	\end{tabular}
		\caption{Calculated energies ($\Delta E$, $E^*_\text{step}$), crossover temperatures ($T_c$) and estimated transition temperatures
	($T_\text{t}$) for the DHT of porphycene at several fcc 110 surfaces.  See definitions in Fig. \ref{fig:Cu110Rates} and text. The  $T_\text{t}$ interval is given by considering calculated or experimental references for the Ag(110) case (see SM). The surface reconstruction of Au(110) \cite{Moritz_SurSci_1979} was ignored for the sake of comparison. \label{tab:Surfaces}}
\end{table}

Using Ag(110) as a reference and building a 1D potential model 
for which it is necessary to calculate the barrier for the concerted mechanism $E^*_\text{con}$ (see model in the SM), $T_\text{t}$ can be estimated for other surfaces.
In Table~\ref{tab:Surfaces} the calculated values for different fcc [110] surfaces are reported, together with the corresponding estimation of $T_\text{t}$.
All $T_c$ values are similar and close to 300 K, showing the importance of tunneling at considerably high temperatures. 
While the estimated $T_t$ represent temperatures that can be achieved in different experiments especially for the stronger interacting surfaces, the resulting rates in Cu(110), Ni(110), and Pd(110) would be smaller than 10$^{-10}$ Hz, which lies far beyond the STM detection limit.

To conclude, we have shown how surface interactions can impact tunneling within a prototype molecular switch based on porphycene molecules adsorbed on metallic surfaces. This study was able to show \rev{that} dimensionality-reduction schemes \rev{for} these problems \rev{would profit from taking into account a temperature dependence in the parametrization}. \rev{The} counter-intuitive origin of different temperature-dependencies of the rates in the deep-tunneling regime and the effects of surface interaction on the \rev{dynamics of} intramolecular hydrogen tunneling \rev{were also resolved}. 
Even though full-dimensional calculations are required to get quantitative results and understand the underlying processes, we could propose a simple estimator to predict the DHT temperature dependence on different metallic surfaces. 

The methodology we presented can be straightforwardly applied to other molecules on surfaces where the calculation of internal hydrogen transfer rates are sought.  Limitations in the RPI approximation may arise when several local minima of the adsorbate with similar energies are present \cite{Kumagai_JCP_2018}. 

The well-defined system addressed in this work 
allowed us to disentangle and quantify static and 
dynamic effects of the environment (in this case the metallic surface) on quantum hydrogen dynamics. We showed that dynamical effects of the environment can promote hydrogen tunneling. Such a quantification is normally not straightforward in condensed phase or biological systems. In this sense, this work shows how single-crystal substrates can be an ideal playground where cutting-edge theory and experiment can meet to provide a deeper understanding of quantum dynamics in fluctuating environments. These findings will help to address hydrogen dynamics in biology 
~\cite{KlinmanKohen2013} and in functional materials~\cite{Tayi_NatChem_2015}, as well as guide the design and interpretation of future experiments.

\begin{acknowledgments}
The authors thank Stuart Althorpe, Aaron Kelly and Matthias Koch for fruitful discussions, and thank Takashi Kumagai and Jeremy Richardson for numerous discussions and a careful assessment of the manuscript. The authors acknowledge financing from the Max Planck Society, and computer time from the
Max Planck Computing and Data Facility (MPCDF) and the Swiss National Supercomputing Centre (CSCS) under project ID s883.
\end{acknowledgments}

\end{document}



\title{Supporting Information \\ Multidimensional Hydrogen Tunneling in Supported Molecular Switches: The Role of Surface Interactions}

\author{Yair Litman}
\email{litman@fhi-berlin.mpg.de}
\affiliation{%
 Fritz Haber Institute of the Max Planck Society, Faradayweg 4--6, 14195 Berlin, Germany, }%
\affiliation{ 
 Institute for Chemistry and Biochemistry, Freie Universität Berlin, Arnimallee 22, 14195 Berlin, Germany}
\author{Mariana Rossi}%
\email{mariana.rossi@mpsd.mpg.de}
\affiliation{%
 Fritz Haber Institute of the Max Planck Society, Faradayweg 4--6, 14195 Berlin, Germany, }%
\affiliation{MPI for the Structure and Dynamics of Matter, Luruper Chaussee 149, 22761 Hamburg, Germany}

\date{\today}

\maketitle

\section{Electronic Structure Calculations}

Density functional calculations were performed using the FHI-aims program \cite{FHI-AIMS}.
The metal surface models were created using the atomic simulation environment ASE \cite{ASE} and were represented by a slab with a 4$\times$6 surface unit cell including 4 layers and a 70 \AA~vacuum in the direction perpendicular to the surface. We have only held the bottom two layers fixed and explicitly considered all other degrees of freedom.  FHI-aims \textit{light} basis sets and numerical settings were used in the production calculations. Calculations with standard \textit{intermediate} settings were also performed for the Cu and Ag surfaces resulting in deviations smaller than 10 meV.
Convergence results with respect to number of layers and k-point sampling of the reciprocal space for all surfaces that were studied are presented in Tab. \ref{table:Cu(110)}-\ref{table:Ni(110)}.
The settings used in the main text are highlighted in the tables and are in good agreement with previously reported values in Ref. \cite{Koch_JACS_2017} using different exchange correlation functionals (optB86b-vdW, BEEF and vdW-DF-cx).

\begin{table}[H]
\centering

        \begin{tabular}{|c||*{2}{c|}}\hline
        \backslashbox[10mm]{layers}{k-grid}&\makebox[3em]{3x3x1}&\makebox[3em]{6x6x1}\\\hline\hline
            4                              &     \textbf{183}            &    188           \\\hline
            6                              &      194           &     -               \\\hline
            8                              &      188           &     -               \\\hline
            4 (PBE + MBD-nl)      &      175           &    -           \\\hline
            4 (HSE06 \cite{HSE06_mod}+ vdW {\cite{TSsurf}}     &      144           &   -           \\\hline
        \end{tabular}
        \caption{ 
        Convergence of reaction energy, E(\textit{trans})-E(\textit{cis}), for the  Cu(110) surface. Values are expressed in meV. Unless otherwise specified, the PBE \cite{PBE} exchange correlation functional augmented with vdW dispersion corrections \cite{TSsurf} (vdW$^\text{surf}$) was used. MBD-nl refers to the non-local version of the many-body-dispersion method \cite{MBD-nl}, and the exchange correlation functional HSE06 was used with $\omega=0.11 \text{bohr}^{-1}$ \cite{HSE06_mod}. Lattice constant: 3.632\AA. } \label{table:Cu(110)}
\end{table}

\begin{table}[H]
\centering
        \begin{tabular}{|c||*{3}{c|}}\hline
        \backslashbox[10mm]{layers}{k-grid}&\makebox[3em]{1x1x1}&\makebox[3em]{3x3x1}&\makebox[3em]{6x6x1}\\\hline\hline
            4                                & \textbf{24}  &       24           &   -            \\\hline
            6                                & -  &       31           &     -               \\\hline
            8                                & -  &       34           &     -               \\\hline
            4 (PBE + MBD-nl)                 & -  &       24           &    -           \\\hline
        \end{tabular}
        \caption{Same as in Tab. \ref{table:Cu(110)} but for the Ag(110) surface. Lattice constant: 4.152\AA.}
\end{table}

\begin{table}[H]
\centering
        \begin{tabular}{|c||*{2}{c|}}\hline
        \backslashbox[10mm]{layers}{k-grid}&\makebox[3em]{3x3x1}   &\makebox[3em]{6x6x1}\\\hline\hline
            4                                 & \textbf{356 / 660}  &       185 / 313        \\\hline
            8                                 & 346 / 646 &         -              \\\hline
 4 (HSE06   +vdW$^\text{surf}$)              & 380 /704       &         -              \\\hline
        \end{tabular}
        \caption{  Energy barriers  for stepwise / concerted mechanism (respectively) at the Cu(110) surface with different simulation settings. Energies expressed in meV. Values reported  for HSE06+vdW$^\text{surf}$ were calculated at PBE+vdW$^\text{surf}$ geometries.}\label{tab:BarrierCu}
\end{table}

\begin{table}[H]
\centering
        \begin{tabular}{|c||*{3}{c|}}\hline
        \backslashbox[10mm]{layers}{k-grid}&\makebox[3em]{1x1x1}&\makebox[3em]{2x2x1}&\makebox[3em]{6x6x1}\\\hline\hline
            4                                & \textbf{188 / 313}  &       185 / 313           &      -         \\\hline
            8                                & 196 / 342  &         -         &     -         \\\hline
             4 (HSE06   +vdW$^\text{surf}$)  & 250 / 455  &         -         &     -         \\\hline
        \end{tabular}
        \caption{Same as in Table \ref{tab:BarrierCu}, but for the Ag(110) surface. \label{tab:BarrierAg} }
\end{table}

\begin{table}[H]
\centering
        \begin{tabular}{|c||*{2}{c|}}\hline
        \backslashbox[10mm]{layers}{k-grid}&\makebox[3em]{3x3x1}&\makebox[3em]{6x6x1}\\\hline\hline
            4                              &       \textbf{72}          &  75              \\\hline
            6                              &       80                   &     -               \\\hline
        \end{tabular}
        \caption{ Convergence of reaction energy, E(\textit{trans})-E(\textit{cis}) for Au(110) surface. Lattice constant: 4.157\AA. See caption in Tab. \ref{table:Cu(110)}. }
\end{table}

\begin{table}[H]
\centering
        \begin{tabular}{|c||*{2}{c|}}\hline
        \backslashbox[10mm]{layers}{k-grid}&\makebox[3em]{3x3x1}&\makebox[3em]{6x6x1}\\\hline\hline
            4                              &       \textbf{151}          &   147              \\\hline
            6                              &       144          &     -               \\\hline
        \end{tabular}
        \caption{ Convergence of reaction energy, E(\textit{trans})-E(\textit{cis}) for Pd(110) surface. Lattice constant: 3.951\AA.  See caption in Tab. \ref{table:Cu(110)}.}
\end{table}


\begin{table}[ht]
\centering
\begin{tabular}{|c||*{2}{c|}}\hline
        \backslashbox[10mm]{layers}{k-grid}&\makebox[3em]{3x3x1}&\makebox[3em]{6x6x1}\\\hline\hline
            4                              &       \textbf{209}         & 210     \\\hline
            6                              &       220        &     -               \\\hline
        \end{tabular}
        \caption{Convergence of reaction energy, E(\textit{trans})-E(\textit{cis}) for  Ni(110) surface. Lattice constant: 3.501 \AA. See caption in Tab. \ref{table:Cu(110)}.}\label{table:Ni(110)}
\end{table}

\section{Reaction and Barrier Energies}

In Tab. \ref{tab:surfaces}, the \textit{cis}$\rightarrow$ \textit{trans} energy difference ($\Delta \tilde{E}$),
the energy barrier of the stepwise mechanism ($\tilde{E}^*_\text{step}$) and the energy barrier of the concerted mechanism 
($\tilde{E}^*_\text{con}$) 
for different metallic surfaces are presented. Note that, differently to the values reported in Tab. II in the main text, $\Delta E$ and $E^*_\text{step}$, the reported values in  \ref{tab:surfaces} are not zero-point energy corrected.

\begin{table}[H]
    \centering
	\begin{tabular}{c|c|c|c}
	\hline
	Surface  & $\Delta \tilde{E}$ & $\tilde{E}^*_\text{step}$ & $\tilde{E}^*_\text{con}$   \\
	\hline
	\hline
    Ag(110)  &  24             &   188     &    313       \\
    Au(110)  &  72              &   216    &    386       \\
    Cu(110)  &  183            &   356     &    660       \\
    Ni(110)  &  180            &   498     &    914       \\ 
    Pd(110)  &  151            &   480     &    822        \\ 
    \hline
	\end{tabular}
		\caption{Calculated reaction  and barrier energies for the DHT of porphycene at several fcc [110] surfaces. Values are not zero-point energy corrected. See definitions in text.  Energies  are expressed meV.}\label{tab:surfaces}
\end{table}

\section{Ring Polymer Instanton Calculations}

The ring polymer instanton (RPI) simulations were performed using the implementation created by us in the i-PI  package \cite{i-pi,i-pi2}. 
Several enhancements in the implementation used for these calculations are available in the open-source code repository and described in Ref. \cite{litman_thesis}.

The force convergence criterion for the transition state geometry and instanton optimizations was 0.005 eV/\AA.
The calculations need to be converged with respect to the number of discretization points of the closed closed Feynman path (CFP), also known as beads.
The high-dimensional neural network reported in  Ref. \cite{Litman_FD_2019} which
delivers an accurate description of the gas phase porphycene potential energy surface was used to perform such convergence tests due to its reduced computational cost. Since it presents a similar vibrational spectrum to the systems considered here, it shows a comparable convergence behaviour. In Tab. \ref{tab:Conv1} and Tab. \ref{tab:Conv2} the convergence of the hydrogen transfer rates for concerted and stepwise mechanisms, respectively, is presented. The reported rates for the DHT of porphycene on Cu(110) and Ag(110) in Fig. 2 in the main text were calculated with a different number of beads at each temperature, in such way to ensure that the error was always below 20\% for the stepwise mechanism and within the order of magnitude for the concerted one. 

\begin{table}[H]
\centering
        \begin{tabular}{|c||*{4}{c|}}\hline
\backslashbox[10mm]{T(K)}{ Beads}&\makebox[3em]{128}&\makebox[3em]{256}&\makebox[3em]{512}&\makebox[3em]{1024}\\\hline\hline
            90       &    \textbf{0.91}  &     0.98       &  1.00  &     -       \\ \hline
            60       &    0.77  &    \textbf{ 0.95}       &  1.00  &     -       \\ \hline
            30       &    0.17  &     0.62       & \textbf{ 0.91}  &  1.00       \\ \hline
        \end{tabular}
    \caption{  Convergence of the reaction rate for the stepwise reaction. For each temperature a value of one is assigned to the converged rate.
    The number of beads used at each temperature for the instanton rate calculations  in this work is highlighted.}
	\label{tab:Conv1} 
\end{table}

\begin{table}[H]
\centering
        \begin{tabular}{|c||*{5}{c|}}\hline
\backslashbox[10mm]{T(K)}{ Beads}&\makebox[3em]{128}&\makebox[3em]{256}&\makebox[3em]{512}&\makebox[3em]{1024}&\makebox[3em]{2048}\\\hline\hline
            60       &  \textbf{0.82}     &  0.96      &  1.00     &    -    & -    \\\hline
            30       &  0.53     & \textbf{0.79}      &  0.92     & 1.00    & -   \\\hline
            10       &  -        &   -        & \textbf{ 0.38}     & 0.69    & 1.00   \\\hline
        \end{tabular}
    \caption{  Convergence of the reaction rate for the concerted reaction. For each temperature a value of one is assigned to the calculation with a bigger amount of beads.
    The number of beads used at each temperature for the instanton rate calculations  in this work is highlighted.}
	\label{tab:Conv2} 
\end{table}




\section{Multidimensional Tunneling Pathways}

\begin{figure*}[htbp]
       \includegraphics[width=0.6\textwidth]{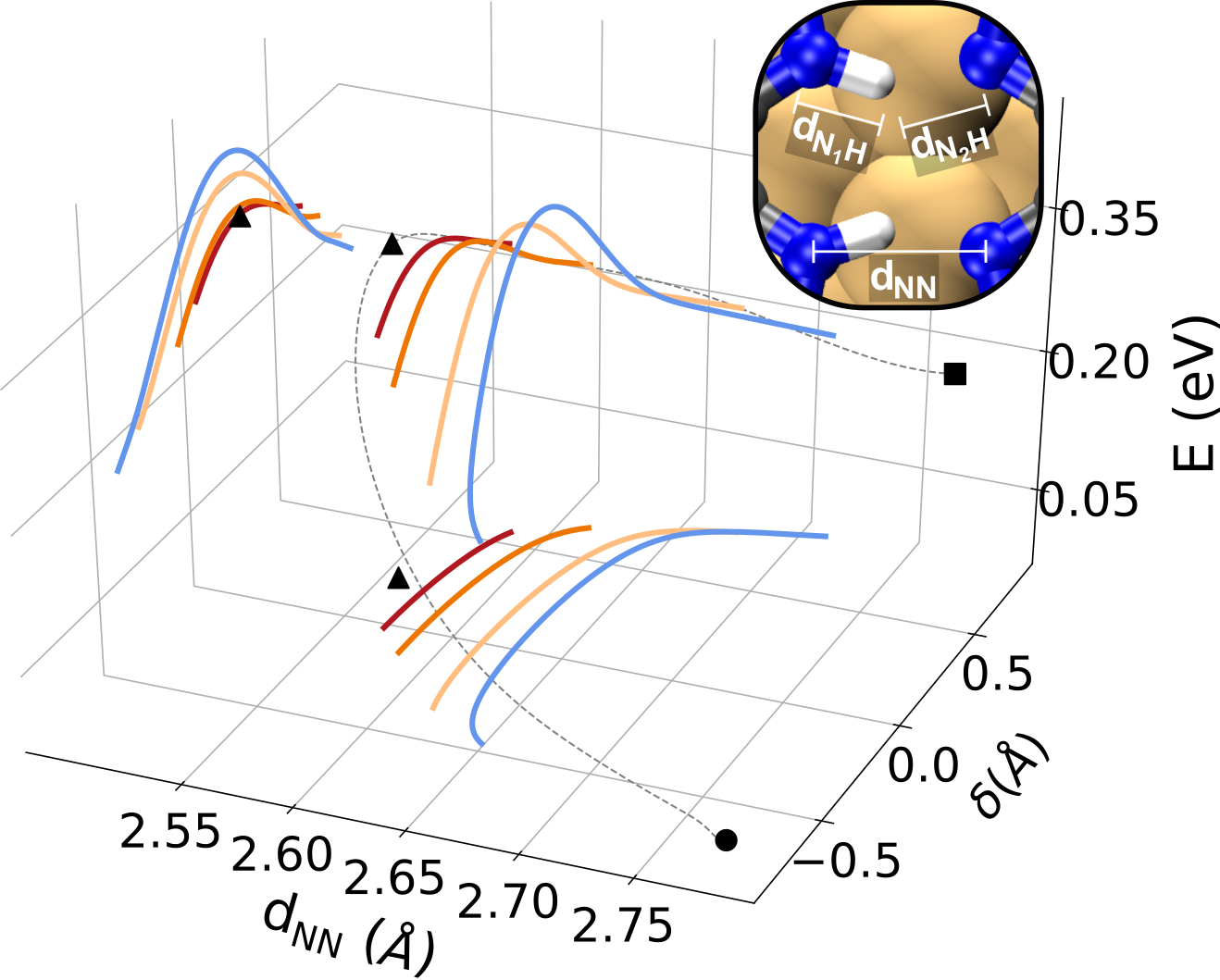}
    \caption{3D visualization of the stepwise instanton pathway at Cu(110) at 40 K (blue), 75 K (orange),
    150 K (dark orange) and 200 K (red). The \textit{cis}, \textit{trans} and transition state geometries are marked with a circle, square and triangle, respectively. Projections are shown with respect to the nitrogen-nitrogen distance $d_\text{NN}$, and the DHT coordinate $\delta = d_{\text{N}_1\text{H}} - d_{\text{N}_2\text{H}}$. The definition guarantees a value of zero when the hydrogen atom is equidistant from $\text{N}_1$ and $\text{N}_2$. 
    The MEP (grey dashed curve) is also plotted as a reference. }
    \label{fig:PathCu}
\end{figure*}

\begin{figure}[H]
\centering
     \includegraphics[width=0.6\columnwidth]{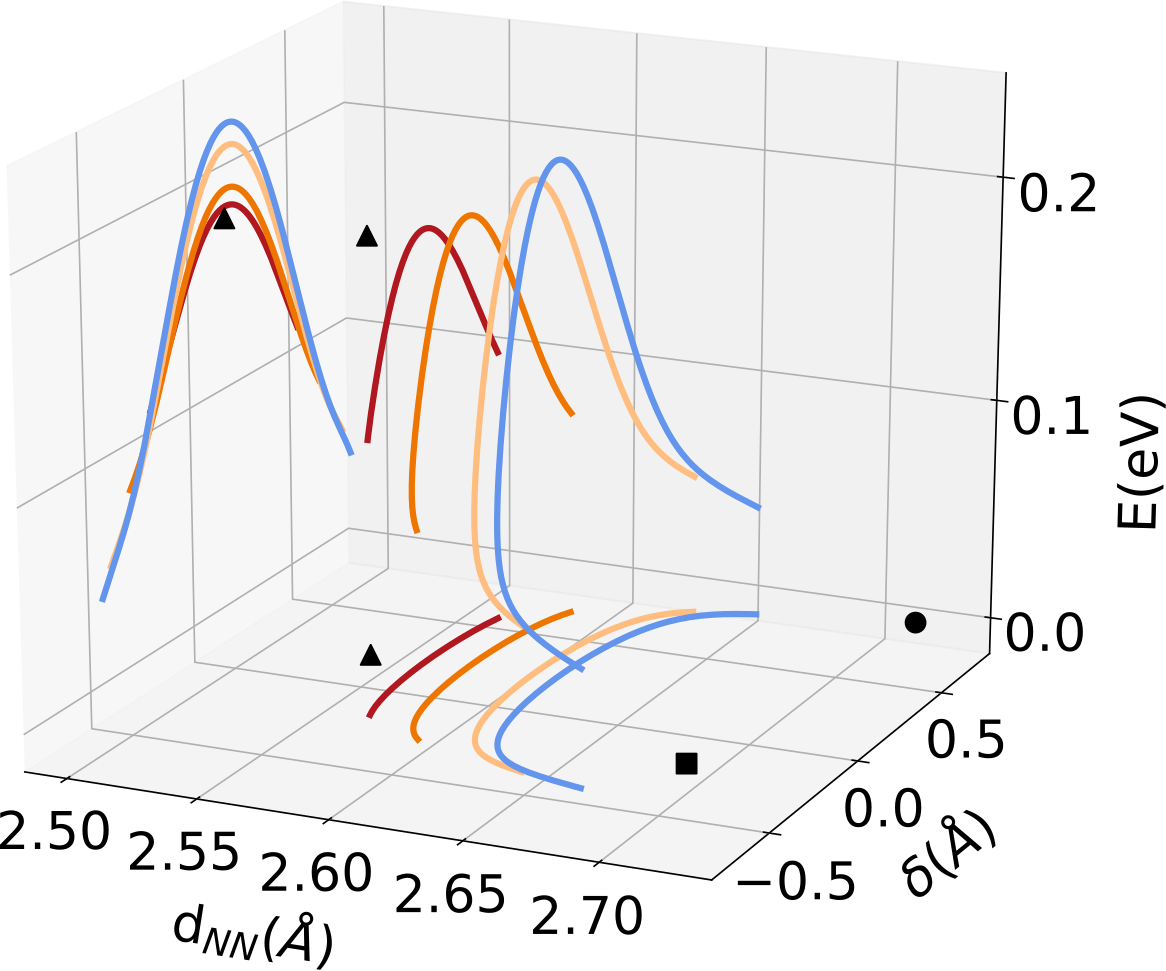}
    \caption{3D visualization of the instanton pathway at Ag(110) at 50K (blue), 80K (orange),
    150K (dark orange) and 200K (red). The \textit{cis}, \textit{trans} and transition state geometries are represented by black circle, square and triangle, respectively. 
    See coordinates definitions in the caption of Fig. \ref{fig:PathCu}.}\label{fig:PathAg}
\end{figure}

In Fig.~\ref{fig:PathCu}, 3D and 2D projections of the minimum energy pathway (MEP) and the instanton tunneling paths at several temperatures on Cu(110) are presented. The MEP involves a shortening of 0.22 \AA~of the distance between nitrogen atoms ($d_{\text{NN}}$) and a flattening of the molecular conformation, as quantified in Tab. \ref{tab:Tun}. The tunneling paths can be compared with the position of the transition state in the MEP, marked by a triangle. The projections show that decreasing the temperature causes the tunneling path to depart further from the position of the transition state.  For example, at 40 K $d_{\text{NN}}$ decreases only 0.1 \AA~ in the tunneling path and it passes through a region 75 meV higher in energy than the transition state point.  The shortening of the tunnelling pathway is the well-known concept coined as corner cutting.  We note that static surface effects that cause the buckling of the \textit{cis} and \textit{trans} conformers on Cu(110) result a considerable increase of the tunneling path length and consequent decrease of $k_{\text{inst}}$ with respect to the gas-phase. Similar effects can be observed in Fig.  \ref{fig:PathAg} for the Ag(110) surface.

\section{Details on the calculation of the DHT on Ag(110)\label{sec:Ag}}

As discussed in the main text, on Ag(110) at 8.5K the concerted mechanism becomes dominant over the stepwise one.
The Euclidean action can be written as $S =\beta \hbar U^\text{RP}$, where $U^\text{RP}$ is the potential energy of 
the ring polymer which represents the discretization of the closed Feynman path \cite{Jor_review_2018}.
 The exponential dependence that the rate has on $S[\mathbf{x}_\text{inst}(\beta)]$ results that at low temperatures (high $\beta$ values) a small error in $S$ has a huge impact on the rate. For example, an error of only 15 meV produces an error in the rate of $\approx e^{17}$ at 10 K, while at 100 K the error is reduced to  only a factor of 5. Thus, while the results for Cu(110) are semi-quantitative, the ones corresponding to the Ag(110) surface at low temperatures should be interpreted only qualitatively.

The  \textit{cis} $\rightarrow$ \textit{cis} reaction in the gas phase and on surfaces only admit stationary trajectories that represent second order saddle points rather than first order ones. In a similar spirit to what has been reported in Ref. \cite{Zhang_PCCP_2014},  the  steepest descent approximation along the second negative mode can be replaced by the full integration in that direction. This procedure was tested on representative 2D systems performing the required numerical integration using 200 points. The results of this procedure were contrasted with taking the module of the second negative frequency and using the standard expression for the rate. While the results differ by one or two orders of magnitude, the temperature dependence is similar (same order of magnitude).  Nevertheless, this effect is minor in comparison to the inaccuracy of several orders of magnitude at low temperatures described before.

\section{Heavy atoms Contribution to the tunnelling process}

In Tab. \ref{tab:Tun}, the tunneling path lengths, computed as the arc length along the instanton pathways,  for the different atomic species at several temperatures are presented.
As expected, the lower the temperature the higher is contribution from heavy atoms to the tunneling dynamics. 
Even though the DHT is much slower at Cu(110) than at Ag(110), the tunnelling path lengths are very similar, showing that the molecule-surface interaction strength do not qualitatively modify the reaction pathway.

\begin{table}[H]
\centering
        \begin{tabular}{|c||*{4}{c|}}\hline
\backslashbox[10mm]{T(K)}{ Species}&\makebox[3em]{H}&\makebox[3em]{C}&\makebox[3em]{N}&\makebox[3em]{Cu}\\\hline\hline
            100      & 0.54/0.58      & 0.07/0.06      &     0.06/0.06      &   0.01/0.00    \\\hline 
            75       & 0.62/0.64  & 0.09/0.09          &  0.09/0.09         &   0.01/0.01    \\\hline 
            50       & 0.73/0.72  & 0.15/0.15         &  0.16/0.15         &  0.03/0.02 \\\hline  
        \end{tabular}
    \caption{ Tunnelling path lengths for H, C, N and Cu atomic species  for the \textit{cis} $\rightarrow$ \textit{trans} reaction at Cu(110) / Ag(110), respectively. Distances are expressed in \AA.}
	\label{tab:Tun} 
\end{table}

\section{Kinetic Isotope Effects on the Cu(110) Surface}

The fact that the kinetic isotope effect (KIE) predicted by transition state theory (TST) is greater than the one predicted by RPI deserves a closer look.
The Eyring TST \cite{Eyring_1935_JCP} is given by

\begin{equation} \label{eq:TST}
\begin{split}
k_\text{TST}(\beta) &=  \frac{1}{2\pi\beta\hbar}\frac{Q^\ddag}{Q_\text{r}}e^{-\beta E^*} =  A_\text{TST}e^{-\beta E^*}, 
\end{split}
\end{equation}

\noindent where $Q^\ddag$ and $Q_\text{r}$ refer to the transition state and reactant partition functions, respectively, $E^*$ to the potential energy barrier, and $\beta=1/k_\text{B}T$ with $k_\text{B}$ the Boltzmann constant and $T$ the temperature.
RPI can be expressed in a similar way, see for example Eq. 1 in the main text and Eq. 8 in Ref. \cite{Beyer_PCL_2016}, where the terms entering the prefactor are explicitly given.
In the main text the activation energy $E_A$ and prefactor refer to the effective Arrhenius slope, $E_A= -\frac{\partial ln(k)}{\partial \beta}$, and effective Arrhenius prefactor, $A =k/e^{-\beta E_A}$. In Table \ref{tab:prefactors}, and in order to allow a rigorous comparison between the considered rate theories, the terms ``exponential factor'' and ``prefactor''  refer, instead, to the factors that appear in Eq.~\ref{eq:TST} above and in Eq. 1 of the main text. 

As discussed already in the main text, the behaviour of the DHT rate in the low temperature limit for the stepwise mechanism presents an effective Arrhenius slope with a value of approximately  $\Delta E$. In the case of the hydrogen isotopologue, $\Delta E$ happens to coincide with $E^*_\text{step}$ while for the deuterium isotopologue $E^*_\text{step}$ is 90 meV larger. 
As a consequence of these energetic relations, and a fortuitous compensation (see Tab. \ref{tab:prefactors}), DHT rates predicted by TST
for the hydrogen isotopologue are unexpectedly similar to the ones predicted by RPI. However, the rates for the deuterium isotopologue are largely underestimated below the cross-over temperature which results in the observed overestimation of KIE.

\begin{table}[h]
\centering
	\begin{tabular}{|c|c|c|c|c|c|c|c|}
	\hline
	$T$ (K)      & $A_\text{inst}$(Hz)  & $e^{-S_\text{inst}/\hbar}$  & $A_\text{TST}$ (Hz) & $e^{-\beta \tilde{E}^*_\text{step}}$  \\
	\hline 
	75           &     $6\times10^{17}$  & $1\times10^{-20}$    & $3\times10^{21}$       & $1\times10^{-24}$\\
	85           &     $1\times10^{17}$  & $2\times10^{-18}$    & $4\times10^{20}$       & $8\times10^{-22}$\\
	100          &     $7\times10^{16}$  & $3\times10^{-16}$    & $2\times10^{19}$       & $1\times10^{-18}$\\
	\hline
	\end{tabular}
	\caption{Dimensionless exponential factor ($e^{-S_\text{inst}/\hbar}$ and $e^{-\beta \tilde{E}^*_\text{step}}$) and prefactors ($A_\text{inst}$ and $A_\text{TST}$) obtained from RPI and TST simulations, respectively. $\tilde{E}^*_\text{step}$ represents the energy barrier for the stepwise mechanism without zero-point energy corrections, because the exponential factor including these corrections is formally in $A_\text{TST}$. Interestingly, even though both theories predict similar DHT rates in this temperature range, the corresponding factors differ significantly.
	 \label{tab:prefactors}}
\end{table}

\section{Estimation of Transition Temperature from Stepwise to Concerted Mechanism in the Deep Tunneling regime}

The low temperature behaviour of the DHT rate for the stepwise mechanism  can be expressed by an Arrhenius-like formula 
\begin{equation} \label{eq:delta}
\begin{split}
k^{T\to 0}_\text{step}(\beta) \approx A_\text{step} e^{-\beta E_A},
\end{split}
\end{equation}
where the zero-point energy corrected reaction energy, ($\Delta E$ in Tab. II.), is a good approximation of the effective activation energy, $E_A$, and the prefactor, $A_\text{step}$, can be approximated by the values obtained from the instanton calculations performed on the Ag(110) and Cu(110) surfaces. On the contrary, the plateau at the low temperature limit  of the rate for the concerted mechanism is difficult to obtain (see Sec. \ref{sec:Ag}). Since this reaction is symmetric, a rough estimate of the rate can be obtained from its tunneling splitting, $\Delta$, as

\begin{equation} \label{eq:conc}
\begin{split}
k^{T\to 0}_\text{con} \approx \frac{\Delta}{2\pi}.
\end{split}
\end{equation}

A reasonable 1D approximation of the potential energy along the reactive coordinate can be written as,  

\begin{equation} \label{eq:1D}
\begin{split}
V(x)=V_0\left[1 - \left(\frac{x}{x_0}\right)^2 \right]^2,
\end{split}
\end{equation}

\noindent where $V_0$ is taken as the  reaction energy barrier ($\tilde{E}^*_\text{con}$ in Tab. \ref{tab:surfaces}), and $x_0$ is a measure of the barrier width that has to be determined.

In Tab. \ref{tab:buclking_cis} and \ref{tab:CC_geo}, the molecular distortions  $h_{\text{bu}}$, measured as the difference between the height of the amino and imino N atoms in the molecule, are presented for the \textit{cis} tautomer (reactant) and the transition state geometry corresponding to the concerted reaction at several surfaces.
At all surfaces considered, the molecule has to reach a flat conformation, characterized by  $h_\text{bu}=0$, for the reaction to take place.
$x^\text{Ag}_0=1.2 $ \AA~and 
$x^\text{Ag}_0=3.0$ \AA~reproduce $\Delta$ values which, by the application of Eq. \ref{eq:delta}, deliver the experimental   and the theoretical plateau rates,  respectively.
The parameters for other surfaces  were obtained by scaling  $x^\text{Ag}_0$  with the ratio of the $h_\text{bu}$ values reported in Tab. \ref{tab:buclking_cis}, i.e. $x^\text{X}_0 = x^\text{Ag}_0 \times (h^\text{X}_\text{bu}/h^\text{Ag}_\text{bu}$). Other scaling factors based on the mass-scaled displacement between the \textit{cis} adsorption conformation and a flat conformation were also tested with similar results. 
\begin{table}[h]
\centering
	\begin{tabular}{|c|c|c|c|c|c|c|c|}
	\hline
	System      & $h_\text{im}$ & $h_\text{am}$ & $h_\text{bu}$    \\
	\hline 
	Ag(110)     &      2.11    &    2.55      &  0.44     \\
	Au(110)     &      2.02    &    2.54      &  0.52     \\
	Cu(110)     &      1.79    &    2.38      &  0.59     \\
	Pd(110)     &      1.68    &    2.25      &  0.56     \\
	Ni(110)     &      1.60    &    2.20      &  0.60     \\
    \hline
	\end{tabular}
	\caption{Average height of nitrogen atoms belonging to the amino groups  ($h_\text{am}$) and imino groups ($h_\text{im}$) for the \textit{cis} tautomer. $h_\text{bu}=h_\text{am}-h_\text{im}$ is a measure of the molecular distortion (buckling) \cite{Kumagai_JCP_2018}. Distances are expressed in \AA. \label{tab:buclking_cis}	}
\end{table}

\begin{table} [h]
\centering
	\begin{tabular}{|c|c|c|c|c|c|c|c|}
	\hline
	System      & $h_\text{im}$ & $h_\text{am}$ & $h_\text{bu}$      \\
	\hline 
	Ag(110)    &      2.32   &    2.32         &  0.0          \\ 
	Au(110)    &      2.34   &    2.34         &  0.0          \\ 
	Cu(110)    &      2.13   &    2.13         &  0.0          \\ 
	Pd(110)    &      2.06   &    2.06         &  0.0         \\
    Ni(110)    &      1.86   &    1.86         &  0.0          \\ 
    \hline
	\end{tabular}
	\caption{Values corresponding to the transition state geometry that connects the two \textit{cis} tautomers. See definitions in Tab. \ref{tab:buclking_cis}.} \label{tab:CC_geo}
\end{table}

Finally, the tunnelling splittings were computed using the Wentzel-Kramers-Brillouin (WKB) approximation \cite{Garg_AmPhys_2000}, and the transition temperature $T_t$ is derived by setting $k^{T\to 0}_\text{con}=k^{T\to 0}_\text{step}$. Both values of $T_t$, calculated from the experimentally derived parameters and from the theoretically derived parameters,  are reported in Tab. II in the main text, giving the reader an assessment of the uncertainty of the estimation. The lower (higher) temperatures correspond to the calculation where the theoretical (experimental) Ag(110) rate  value was used as a reference.